\documentclass[twocolumn,english,aps,reprint, superscriptaddress,showpacs,longbibliography, showkeys,nofootinbib]{revtex4-2}
\usepackage{amsmath,amssymb,bbm,mathrsfs,bm,braket,color,graphicx,comment,xcolor} 
\usepackage[colorlinks,citecolor=blue,urlcolor=blue,linkcolor = blue]{hyperref}
\usepackage[mathscr]{euscript}
\usepackage{hyperref}
\usepackage{enumitem}

\begin{document}

\title{Modular Parity Quantum Approximate Optimization}

\author{Kilian Ender}
\affiliation{Institute for Theoretical Physics, University of Innsbruck, A-6020 Innsbruck, Austria}
\affiliation{Parity Quantum Computing GmbH, A-6020 Innsbruck, Austria}

\author{Anette Messinger}
\affiliation{Parity Quantum Computing GmbH, A-6020 Innsbruck, Austria}

\author{Michael Fellner}
\affiliation{Institute for Theoretical Physics, University of Innsbruck, A-6020 Innsbruck, Austria}
\affiliation{Parity Quantum Computing GmbH, A-6020 Innsbruck, Austria}

\author{Clemens Dlaska}
\affiliation{Institute for Theoretical Physics, University of Innsbruck, A-6020 Innsbruck, Austria}
\affiliation{Institute for Quantum Optics and Quantum Information of the Austrian Academy of Sciences, A-6020 Innsbruck, Austria}

\author{Wolfgang Lechner}
\affiliation{Institute for Theoretical Physics, University of Innsbruck, A-6020 Innsbruck, Austria}
\affiliation{Parity Quantum Computing GmbH, A-6020 Innsbruck, Austria}

\date{\today}

\begin{abstract}
The parity transformation encodes spin models in the low-energy subspace of a larger Hilbert-space with constraints on a planar lattice. 
Applying the Quantum Approximate Optimization Algorithm (QAOA), the constraints can either be enforced explicitly, by energy penalties, or implicitly, by restricting the dynamics to the low-energy subspace via the driver Hamiltonian. 
While the explicit approach allows for parallelization with a system-size-independent circuit depth, the implicit approach shows better QAOA performance. Here we combine the two approaches in order to improve the QAOA performance while keeping the circuit parallelizable. In particular, we introduce a modular parallelization method that partitions the circuit into clusters of subcircuits with fixed maximal circuit depth, relevant for scaling up to large system sizes. 
\end{abstract}

\maketitle

\section{Introduction}
As quantum technology advances \cite{Saffman2010, childress2006coherent, obrien2009, Blatt2012, kokail2019, Kjaergaard2020, hollerith2021, Scholl2021, Ebadi2021}, there is a large ongoing effort to apply quantum algorithms to optimization problems \cite{Hauke2020, bharti2021noisy}, with the goal of achieving a quantum computational advantage \cite{Arute2019, Zhong2020, Wu2021}. A gate-based algorithm designed for solving combinatorial optimization problems on contemporary noisy quantum devices is the Quantum Approximate Optimization Algorithm (QAOA) \cite{Farhi2014, Farhi2016, Crooks2018,Zhou2020, Hastings2019}. First proof-of-principle QAOA-implementations for specific problem graphs have already been successfully demonstrated on quantum hardware \cite{Pagano2020, Harrigan2021, Ebadi2022}.
However, hardware implementations of QAOA for generic combinatorial optimization problems are challenging due to the limited inter-qubit connectivity of quantum devices, which can be detrimental for practical QAOA performance \cite{Harrigan2021}. 

The recently introduced parity architecture \cite{Lechner2015, Ender2021} addresses this mismatch between the connectivity of the problem and hardware graphs by mapping problem-defining interactions onto single-body terms, while restricting the enlarged Hilbert space via quasilocal constraint terms.
In particular, the parity architecture allows one to tackle generic optimization problems, i.e. problems with\ long-range and higher-order couplings, on a problem-independent and fixed qubit layout utilizing only quasilocal interactions.

In previous works \cite{lechner2020, Dlaska2021}, QAOA implementations for the parity architecture have been proposed where the constraints are \textit{explicitly} enforced through an energy penalty. However, it has been shown that preserving the constraint conditions can be achieved \textit{implicitly} by making the involved operators commute with the constraint operators \cite{Rocchetto2016_stabilizer, hen2016_constrained_drivers, hadfield2019_quantumalternating}. While the former enables a parallelizable implementation with low circuit depth in the QAOA, the latter can lead to significantly improved success probabilities.

\begin{figure}[t]
    \centering
    \includegraphics[width=\columnwidth]{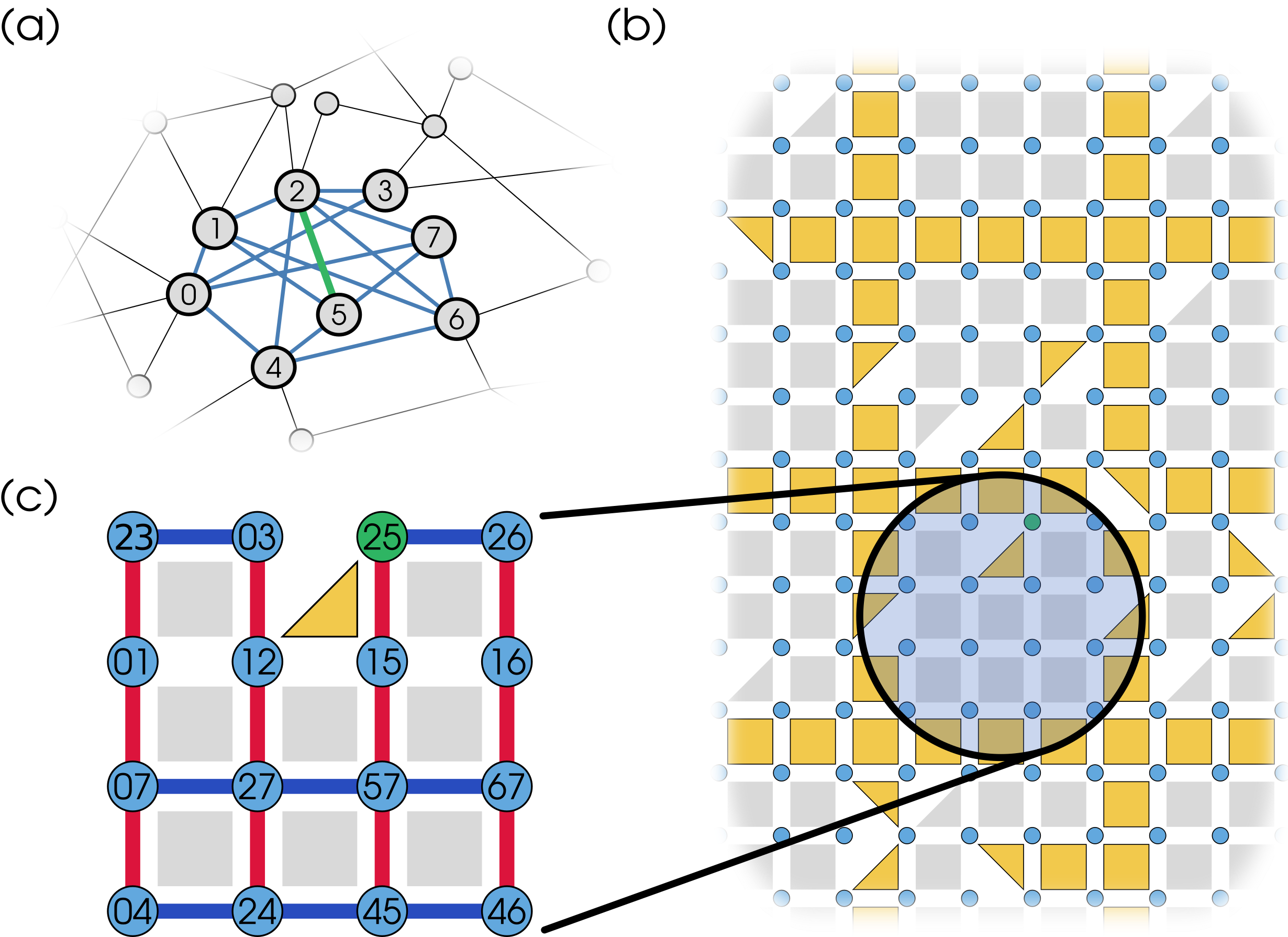}
    \caption{Modularization of a parity-compiled problem. (a) Problem graph to be implemented, with a subgraph highlighted in blue. (b) Implementation layout of the parity encoded problem. Blue dots represent parity qubits, gray (yellow) squares and triangles represent implicitly preserved (explicitly enforced) three- and four-body constraints. Explicitly enforced constraints are used to divide the layout into modules. (c) One module of the layout, corresponding to the highlighted subgraph in (a), with terms of the driver Hamiltonian illustrated by red and blue lines. The modularization leads to a highly parallelizable circuit implementation of the required driver terms. The green highlighting illustrates the correspondence of interactions in (a) to parity qubits in (b) and (c).
    }
    \label{fig:fig1}
\end{figure}

In this work we suggest a \textit{hybrid} approach, which keeps the required circuit depth constant while reducing the number of constraints to be enforced explicitly and thus improving performance. We do this by partitioning the constraints into a set that is enforced explicitly and a set where the constraints are preserved implicitly by adapting the driver Hamiltonians. Fig.~\ref{fig:fig1}c shows an example layout of encoded qubits and constraints for a problem described by the subgraph highlighted in blue in Fig.~\ref{fig:fig1}a.
In this layout a single constraint is enforced explicitly (yellow) while the others are implicitly preserved by the driver, which acts on qubits in each of the shown lines simultaneously. By choosing which constraints are in which set, we can divide bigger layouts into smaller modules (see Fig.~\ref{fig:fig1}b), enabling a parallel implementation of all required unitaries with an adjustable maximal circuit depth.

\section{Parity QAOA}
Finding solutions to generic combinatorial optimization problems can be formulated as energy minimization of general (classical) $N$-spin Hamiltonians of the form
\begin{equation}\label{eq:problem_hamiltonian}
\begin{split}
    H_\text{problem} = & \sum_{i} J_{i} s_i +  \sum_{i<j} J_{ij} s_i s_j 
    \\& + \sum_{i<j<k} J_{ijk} s_i s_j s_k + \dots,
\end{split} 
\end{equation}
where ${s_i=\pm 1}$ denote spin variables and the coefficients $\lbrace J_i, J_{ij}, J_{ijk},\dots\rbrace$ describe long-range and potentially higher-order interactions between spins. We denote the number of non-zero coefficients by $K$ and the number of spin-flip symmetries in the Hamiltonian by $n_\text{s}$. Note that we do not consider problems with side-conditions in this work.

On state-of-the-art hardware platforms, the available inter-qubit couplings are typically two-body and limited in distance. Therefore, interactions as occurring in Eq.~\eqref{eq:problem_hamiltonian} can be challenging to implement directly. Instead, we utilize the parity architecture \cite{Lechner2015, Ender2021} that allows one to encode arbitrary $k$-body terms on a square lattice requiring only nearest-neighbor interactions.
This involves mapping the product of $k$ problem spins $s_i$ onto a single, physical parity qubit (denoted by $\hat\sigma_z$), e.g.\@ $ J_{ijk}\, s_i s_j s_k \mapsto J_m\, {\hat\sigma}^{(m)}_z $, where we label each parity qubit with the corresponding $k$-tuple $m$ of problem spin indices. 
As a result, the $K$ problem-defining interaction terms of Eq.~\eqref{eq:problem_hamiltonian} are represented by local fields of strength $J_m$ acting on $K\geq N$ parity qubits. This gives rise to a physical Hamiltonian of the form $\hat H_\text{phys} = \hat H_\text{Z} + \hat H_\text{C}$, where  
\begin{equation}\label{eq:localfield_hamiltonian}
    \hat H_\text{Z} = \sum_{m} J_{m}\hat\sigma_z^{(m)}
\end{equation}
encodes the combinatorial optimization problem and $\hat H_\text{C}$ contains constraints to ensure that the code space corresponds to the low-energy subspace of the enlarged Hilbert space $\mathcal{H}_\text{phys}$. The constraint Hamiltonian $\hat H_\text{C}$ is constructed as
\begin{equation}\label{eq:constraint_ham}
 \hat H_\text{C} = \sum_l \hat{C}_l
\end{equation}
with three- or four-body interactions (the square brackets indicate the optional factor)
\begin{equation}\label{eq:constraint_term}
\hat{C}_l = \frac{c_l}{2} \left(1-\hat\sigma_z^{(l_1)}\hat\sigma_z^{(l_2)}\hat\sigma_z^{(l_3)}\,[\hat\sigma_z^{(l_4)}]\right)
\end{equation}
acting on $2\times2$ plaquettes of physical qubits (cf.\ Fig.~\ref{fig:fig1}) and a constraint strength $c_l>0$ \cite{Lanthaler2021}.
Here, $l_i$ are labels of physical qubits with the property that all problem-spin indices involved in constraint $\hat{C}_l$ appear an even amount of times across all the $l_i$. 
Following this construction, constraint-satisfying states are characterized by an even number of qubits in the $\ket{\downarrow}$-state per constraint (with ${\hat\sigma_z \ket{\downarrow}=-\ket{\downarrow}}$), and the code space coincides with the \textit{constraint-fulfilling subspace}
\begin{equation}
    \mathcal{H}_\text{CF} 
    =\left\{\ket{\psi} \in \mathcal{H}_\text{phys} \, \Big| \, \hat H_\text{C}\ket{\psi}=0\right\} \text{.}
\end{equation}
{Figure~\ref{fig:all_to_all}a} shows the parity implementation of an all-to-all connected Ising spin-glass model, where all constraints are explicitly enforced by three- and four-body interactions.

\subsection{Explicit Parity QAOA}
The QAOA \cite{Farhi2014} attempts to find low energy solutions of $H_\text{problem}$ by evolving a quantum state alternately with a driver Hamiltonian $\hat{H}_\text{B} = \sum_{i=1}^N \hat\sigma_x^{(i)}$ and the (quantum mechanical) problem Hamiltonian $\hat H_\text{problem}$ for variable durations. To implement the QAOA in the parity architecture \cite{lechner2020}, the single-qubit driver Hamiltonian $\hat H_\text{X} = \sum_{m} \hat\sigma_x^{(m)}$ now acts on $K$ physical qubits, while the problem Hamiltonian $\hat H_\text{problem}$ is replaced by the two components of ${\hat H_\text{phys}}$, i.e.\ $\hat H_\text{Z}$ and $\hat H_\text{C}$. A parity-QAOA sequence of depth $p$ thus corresponds to variationally evolving the system with Hamiltonians $\hat H_\text{X}$, $\hat H_\text{Z}$ and $\hat H_\text{C}$ as
\begin{equation}\label{eq:qaoa_state_lhz}
\ket{\psi} = \prod_{j=1}^p e^{-i\beta_j \hat{H}_\text{X}}e^{-i\gamma_j \hat{H}_\text{Z}}e^{-i\Omega_j \hat{H}_\text{C}}\ket{+}^{\otimes K},
\end{equation}
where the variational parameters $\beta_j$, $\gamma_j$ and $\Omega_j$ are optimized in a quantum-classical feedback loop in order to minimize $\braket{\psi | \hat H_\text{phys} | \psi}$.
As a consequence, during optimization, the constraints introduced by the parity mapping are treated on the same footing as the problem encoding single-body terms, and the QAOA unitary $e^{-i\Omega \hat{H}_C}$ needs to be implemented \textit{explicitly} in order to steer the dynamics into $\mathcal{H}_\text{CF}$.

\subsection{Implicit Parity QAOA}\label{sec:implicit_parity_qaoa}
\begin{figure*}[ht]
\centering\includegraphics[width=\textwidth]{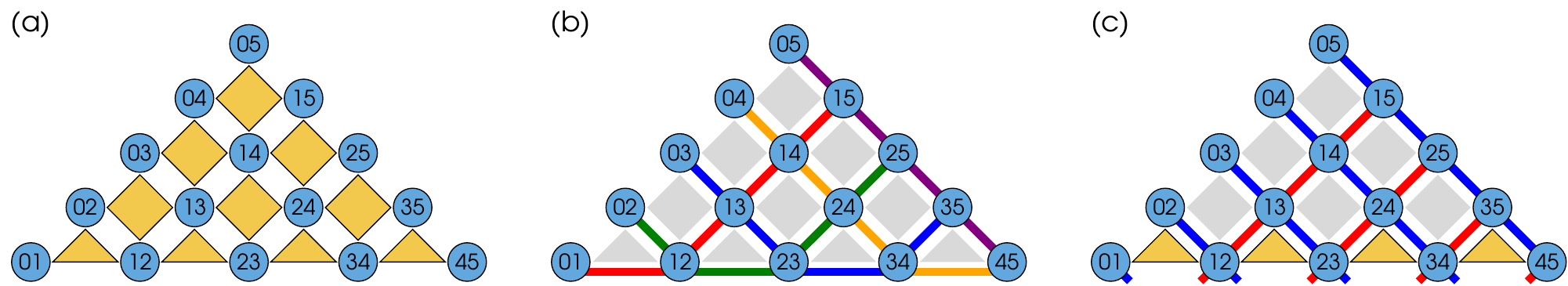}
\caption{Example of a parity-encoded complete graph with six spin variables. (a) All constraints are enforced explicitly via the constraint Hamiltonian. The driver Hamiltonian contains single-qubit $\hat\sigma_x$ operators on all physical qubits. (b) The colored lines denote sets of qubits, each of which can be flipped simultaneously without leaving the constraint-fulfilling subspace (constraint-preserving driver lines). All constraints are implicitly enforced via the driver Hamiltonian, and no energy penalty for constraints is needed. 
(c) Only the bottom-most row of 3-body constraints is enforced explicitly, while the others are satisfied implicitly due to the restriction of dynamics via the driver Hamiltonian. In this setting, the blue and red driver lines (hybrid driver lines) can be implemented in parallel, respectively.}
 \label{fig:all_to_all}
\end{figure*}

An alternative approach to perform parity QAOA is to start with a state in $\mathcal{H}_\text{CF}$ and restrict the dynamics to that subspace by adapting the driver Hamiltonian \cite{hadfield2019_quantumalternating, hen2016_constrained_drivers}. The constraint conditions then are preserved \textit{implicitly} throughout the QAOA sequence, which implies that we are looking for a driver Hamiltonian $\hat{H}_\text{X}^{\text{imp}}$ fulfilling 
\begin{equation}\label{eqn:commutator_logical}
[\hat{H}_\text{C}, \hat{H}_\text{X}^{\text{imp}}]=0 \text{.}
\end{equation}
We choose $\hat{H}_\text{X}^{\text{imp}}$ to be a sum over products of $\hat{\sigma}_x$ operators.
As the constraints are products of $\hat \sigma_z$ operators, each constraint term commutes with individual terms of $\hat{H}_\text{X}^{\text{imp}}$ whenever they share an even number of qubits.
For Eq.~\eqref{eqn:commutator_logical} to hold, each term in $\hat{H}_\text{X}^{\text{imp}}$ must commute with each constraint term independently.

{Fig.~\ref{fig:all_to_all}b} shows the qubits in such terms of $\hat{H}_\text{X}^{\text{imp}}$ as colored lines for the example of an all-to-all connected problem graph \cite{Rocchetto2016_stabilizer}. In this example, the qubit labels contributing to a certain line share a common problem spin index, and therefore can be associated to this particular problem spin. Thus, the sum over all such products is the parity-mapped analogue of the ``standard'' driver Hamiltonian $\hat{H}_\text{B}$ acting on the problem spins.
In the following we formalize the aforementioned considerations and define the elements of constraint-preserving driver Hamiltonians.

First, we consider a set of physical qubits that can be flipped simultaneously without changing which constraint conditions are preserved. These qubits are typically arranged on the layout along a line (see for example the colored lines in Fig.~\ref{fig:all_to_all}b), or in more general cases manifest as a tree graph of adjacent qubits. 
In the following, we refer to these sets as constraint-preserving \textit{driver lines} $Q^\mu$, with the index $\mu$ enumerating the driver lines for a given problem.

With each driver line, we associate a \textit{driver term}
\begin{equation}\label{eq:logical_operators}
\hat X^{(\mu)}=\prod_{k\in Q^\mu} \hat\sigma_x^{(k)}
\end{equation}
with the property
\begin{equation}
\ket{\psi}\in \mathcal{H}_\text{CF} \Longleftrightarrow \hat X^{(\mu)}\ket{\psi}\in \mathcal{H}_\text{CF}\text{.}
\end{equation}
We refer to the number of qubits in a driver line as its \textit{length}.
Our goal is to explore the full code space with a set of driver terms that allows for independent flips of any problem spin, which requires the following properties for driver lines: 
A set $D$ of driver terms is \textit{independent} iff no element ${Q^\mu \in D}$ can be obtained as a product of (multiple) other elements in $D$.
Furthermore, we call the set $D$ \textit{valid} iff $D$ is independent and ${|D|=N-n_\text{s}}$ holds. This ensures that each of the $N-n_\text{s}$ independent spins of the original problem can be flipped.
Two driver lines $Q^\mu$ and $Q^\nu$ are said to \textit{overlap}, iff ${Q^\mu \cap Q^\nu \neq \emptyset}$. In the following, we use $D$ to refer to a set of driver lines as well as to the set of its associated driver terms. Note that the product of driver terms translates to the symmetric difference of associated driver lines.

In contrast to the standard parity-QAOA approach enforcing all constraints explicitly as discussed in \cite{lechner2020}, we now investigate the performance of parity QAOA utilizing driver Hamiltonians
\begin{equation}\label{eq:hx}
    \hat H_\text{X}^{\text{imp}}= \sum_{\mu = 1}^{N-n_\text{s}} \hat X^{(\mu)}\text{,}
\end{equation}
consisting of the operators associated with a valid set of constraint-preserving driver lines. Provided that we start from a constraint-fulfilling state, such a driver only introduces transitions to other constraint-fulfilling states and therefore restricts the dynamics to $\mathcal{H}_\text{CF}$.

Using the constraint-preserving driver Hamiltonian, the corresponding QAOA protocol is given by
\begin{equation}\label{eq:logical_qaoa_protocol}
\ket{\psi} = \prod_{j=1}^p e^{-i\beta_j \hat{H}_\text{X}^\text{imp}}e^{-i\gamma_j \hat{H}_\text{Z}}\ket{\psi_0},
\end{equation}
with $\ket{\psi_0}$ being an appropriately chosen initial state fulfilling all parity constraints.
Usually, $\ket{\psi_0}$ is chosen to be the equal superposition of all constraint-fulfilling computational states (for details on the initialization see Sec. \ref{sec:initial_state_preparation}).
Note that compared to the QAOA-protocol described in Eq.~\eqref{eq:qaoa_state_lhz}, the step involving $\hat{H}_\text{C}$ is not required anymore, since all constraints are now implicitly preserved and do not have to be enforced by the constraint Hamiltonian.
Apart from saving one variational parameter per QAOA-cycle, the intrinsic fulfillment of parity constraints also results in an exponential reduction of the size of the accessible Hilbert space, decreasing the probability of populating undesired states and thus significantly enhances the performance of the algorithm.

Hamiltonians with multi-qubit terms of the form $\hat X^{(\mu)}$ can in general not be simulated in quantum hardware directly. The unitary operator $\hat U=e^{-i\beta \hat X^{(\mu)}}$, however, can be readily implemented as a sequence of CNOT-gates and single qubit rotations \cite{Cowtan2020}, with a circuit depth scaling linear in the length of the driver line $Q^{\mu}$ (see Appendix~\ref{sec:app_decompose}).

\section{Hybrid approach and modularization}
A drawback of the fully implicit QAOA implementation described in Section \ref{sec:implicit_parity_qaoa} is that the driver lines can become arbitrarily long or overlap, which limits the ability to perform gates in parallel to achieve a low overall circuit depth.
In particular, a fully implicit implementation of the complete graph requires a driver unitary with a circuit depth scaling at least linearly with the system size $N$\footnote{In this case, the length of a single driver line and therefore also its implementation depth is already proportional to $N$.}. 
In order to keep the circuit depth feasible, especially for non-error-corrected quantum devices, we introduce a \textit{hybrid} implementation as a way to balance between the advantages of the fully explicit and the fully implicit approaches.
The main idea is to shorten/split driver lines to obtain a parallelizable implementation that requires a minimal number of explicitly enforced constraints (cf.\ Fig.~\ref{fig:all_to_all}c).
This can be achieved by separating the required $n_\text{C}^\text{tot}$ constraints into $n_\text{C}$ explicitly enforced constraints and $n_\text{C}^\text{tot}-n_\text{C}$ implicitly preserved constraints. The resulting hybrid Hilbert space $\mathcal{H}_\text{hyb}$ is thus spanned by the computational basis states fulfilling all implicitly preserved constraints with
\begin{equation}
    \dim(\mathcal{H}_\text{hyb})=2^{N+n_\text{C}-n_\text{s}}.
\end{equation}
Similar to the implicit approach, a \textit{hybrid driver line} $Q^\mu$ is given by a set of physical qubits that can be simultaneously flipped without changing the population in the hybrid subspace $\mathcal{H}_\text{hyb}$. 
The definitions of \textit{length}, \textit{overlap}, \textit{independence} and \textit{driver terms} for the hybrid case are analogously defined as for the implicit approach introduced in Sec.~\ref{sec:implicit_parity_qaoa} w.r.t.\ $\mathcal{H}_\text{hyb}$.

A set $D$ of hybrid driver terms is \textit{valid} iff it is independent and any computational basis state in the constraint-fulfilling Hilbert space $\mathcal{H}_\text{CF}$ can be transformed to any other by applying operators in $D$ only. Note that this requirement is less strict compared to fully constraint-preserving driver lines, since $D$ can contain ${N-n_\text{s} \leq |D| \leq N+n_\text{C}-n_\text{s}}$ driver terms.
In the present work we focus on ${|D| = N+n_\text{C}-n_\text{s}}$, since in all other cases, there are explicitly enforced constraints which are naturally preserved by the driver lines.

As a consequence, for a problem with $N$ spin variables and $n_\text{C}^\text{tot}$ constraints of which $n_\text{C}$ are enforced explicitly by
\begin{equation}
    \hat H_{\rm C}^\text{hyb} = \sum_{l=1}^{n_\text{C}} \hat C_l,
\end{equation} we can choose the \textit{hybrid} driver Hamiltonian as
\begin{equation}
    \hat H_{X}^\text{hyb}=\sum_{\mu=1}^{N+n_\text{C}-n_\text{s}} \hat X^{(\mu)},
\end{equation}
with driver terms $\hat X^{(\mu)}$ associated to a valid set of hybrid driver lines. 
In contrast to the fully implicit implementation, we can no longer associate individual driver terms to single-qubit operations on the original problem spins.
Note that the fully implicit and the fully explicit approach correspond to the limiting cases of the hybrid approach with ${n_\text{C}=0}$ and ${n_\text{C}=n_\text{C}^\text{tot}}$, respectively. 

The QAOA-protocol is now given by 
\begin{equation}\label{eq:qaoa_protocol_hybrid}
\ket{\psi} = \prod_{j=1}^p e^{-i\beta_j \hat H_{X}^\text{hyb}} e^{-i\gamma_j \hat{H}_\text{Z}}e^{-i\Omega_j \hat H_{\rm C}^\text{hyb}} \ket{\psi_0},
\end{equation}
with replacements ${\hat H_{X} \mapsto \hat H_{X}^\text{hyb}}$ and ${\hat H_{\rm C} \mapsto \hat H_{\rm C}^\text{hyb}}$ compared to the protocol described in Eq.~\eqref{eq:qaoa_state_lhz}. The initial state $\ket{\psi_0}$ is typically chosen to be the equal superposition of all computational basis states in $\mathcal{H}_\text{hyb}$ (see Sec.~\ref{sec:initial_state_preparation} for details on the initialization). 

In the following, we demonstrate our hybrid approach on the example of the complete graph and then show how this can be applied to arbitrary graphs. At the end of this section we introduce the concept of modularization in order to extend our approach to large system sizes with system-size independent circuit depths. 

\subsection{Example: Complete graph}
In this section we illustrate the above introduced concepts on the example of a parity-encoded problem graph with all-to-all connectivity as pictured in Fig.~\ref{fig:all_to_all}. Starting from a constraint-fulfilling state, flipping a single physical qubit leads to the violation of at least one constraint. When flipping more qubits until all constraints are fulfilled again, the minimal set of flipped qubits will correspond to a constraint-preserving driver line as in Fig.~\ref{fig:all_to_all}b. The corresponding driver terms, however, cannot be implemented in parallel and result in impractical circuit depths. 

A particular way to render the circuit shorter and parallelizable is to ``break'' each long constraint-preserving driver line into two shorter driver lines by enforcing all three-body constraints explicitly as depicted in Fig.~\ref{fig:all_to_all}c. 
Note that in such a setting, the original driver lines (cf.\ Fig.~\ref{fig:all_to_all}b) also remain valid, even though the bottom constraints are explicitly enforced. 
Switching a constraint from an implicit to an explicit implementation doubles the dimension of the reachable subspace by including states that violate the corresponding explicitly enforced constraint. This increased flexibility allows one to split the original driver line into two shorter driver lines, such that each of these two lines violate the switched constraint and the symmetric difference of the two lines restores the original driver line. 

The resulting driver Hamiltonian can be easily parallelized by classifying the lines into two groups by their orientation in the layout. In Fig.~\ref{fig:all_to_all}c, this classification is represented by the red and blue coloring of lines. As none of the lines within a group overlap, their corresponding gate sequences can be executed at the same time. Hence, the implementation of the total driver unitary $\exp(-i \beta \hat H_\text{X}^{\text{hyb}})$ takes a circuit depth of at most $2N$. This can be further reduced to a constant depth by modularization of the layout, as explained in section \ref{sec:modularization}.

\subsection{Arbitrary (hyper-)graphs}\label{sec:arbitrary_graphs}

\begin{figure}[tb]
\includegraphics[]{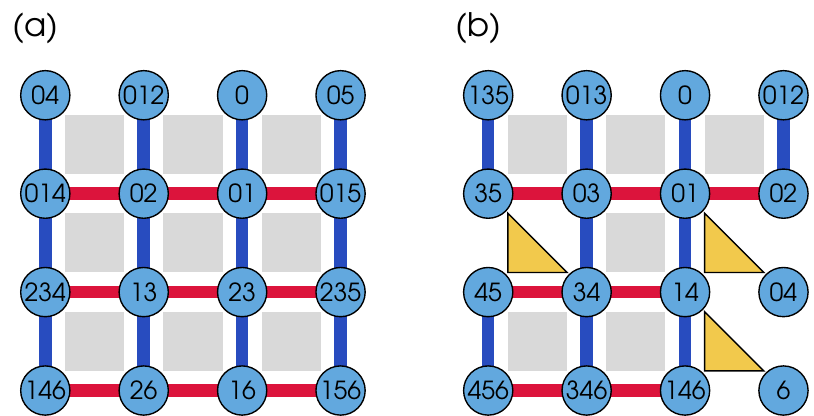}
\caption{Constraint layout examples with partitioning into three- and four-body constraints. (a) The special case with only four-body constraints for which the driver lines may be chosen to be strictly horizontal or vertical and can be parallelized trivially. (b) General example with both, three- and four-body constraints. Implementing all three-body constraints explicitly still allows for parallel execution of all horizontal (vertical) lines. The shown driver terms preserve all four-body constraints (gray), while three-body constraints have to be explicitly enforced in $\hat H_\text{C}$ (yellow). The driver line in the first row has been omitted as it can be obtained via symmetric difference of the others. Qubits not involved in any of the shown driver lines are part of a single-qubit driver (not depicted).}
\label{fig:mixed}
\end{figure}

Compiling more general graphs, and in particular hypergraphs, to the parity architecture leads to a variety of placements of three- and four-body constraints among qubits in a square lattice geometry  (cf.\ Fig.~\ref{fig:mixed} and Ref.~\cite{Ender2021}). 
In the simplest case, requiring only four-body constraints, we can construct a driver Hamiltonian which preserves all constraints from only straight horizontal and vertical lines (cf. Fig~\ref{fig:mixed}a). This is still true for most layouts with mixed three- and four-body constraints where all three-body constraints are enforced explicitly (see Fig.~\ref{fig:mixed}b). 
The only exception are layouts with isolated groups of three-body constraints, which are not connected to the boundary of the layout through adjacent explicitly enforced constraints. Enforcing isolated constraints explicitly can require more complicated driver lines, including turns and branches. This can be circumvented by explicitly enforcing additional constraints until all isolated explicitly enforced constraints are connected to the boundary via other explicitly enforced constraints. 
Hence, a simple strategy to partition the constraints is to  explicitly enforce all three-body constraints, and all four-body constraints required to connect them to the boundary, while the remaining four-body constraints are implicitly preserved by the drivers. The full driver circuit can then be implemented in two steps, where all horizontal and all vertical driver lines are implemented in parallel, respectively.

\begin{figure}[tb]
\includegraphics[]{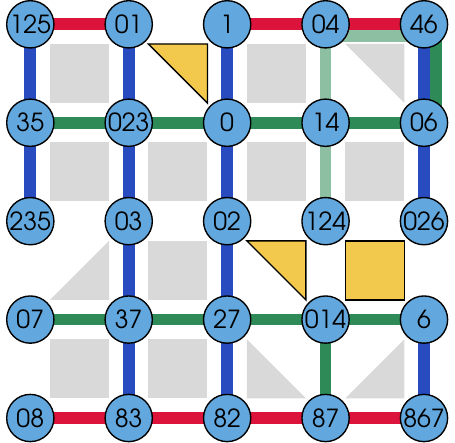}
\caption{Example of an optimized set of explicitly enforced constraints (yellow). The optimization aims at increasing the number of implicitly preserved three-body constraints (gray triangles) which cause the driver lines shown in green to deviate from straight line shapes. The yellow square is a four-body constraint which is kept explicitly enforced to connect the adjacent explicitly enforced three-body constraint to the boundary, simplifying the driver lines. One line has been omitted as it can be obtained via symmetric difference of the others.
}
\label{fig:optimized}
\end{figure}

In many cases, the number of of explicitly enforced constraints can be further reduced since some of the three-body constraints are automatically preserved by the above mentioned horizontal and vertical driver lines.
In other cases, small adjustments to the driver lines by adding or removing qubits (which may introduce turns and branches), are sufficient to preserve even more three-body constraints. An example of minimizing the number of explicitly enforced constraints at a fixed maximal circuit depth can be seen in Fig.~\ref{fig:optimized}.

\subsection{Modularization}\label{sec:modularization}
With the procedure described in the previous section, the average length of hybrid driver lines (and therefore the depth of the QAOA-circuit) grows linearly with the dimensions of the layout. We now utilize the concept of implicitly preserved and explicitly enforced constraints to restrict the driver circuit depth to an adjustable and system-size independent value, while minimizing the number of explicitly enforced constraints. 

\begin{figure}[tb]
\includegraphics[angle = 90, width=0.97\columnwidth]{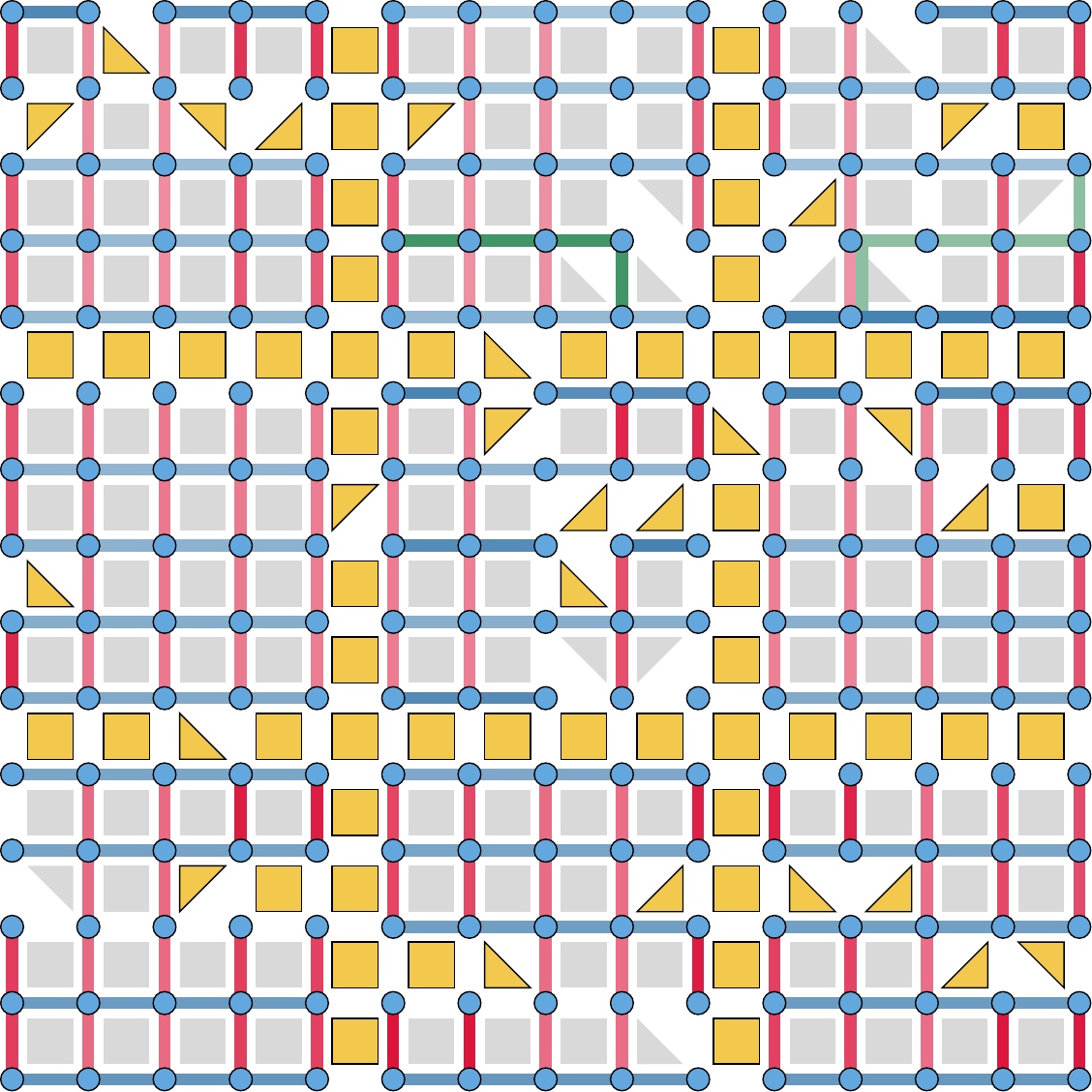}
\caption{Modularization of a larger layout with additional explicitly enforced constraints (yellow) arranged in a grid for constant circuit depth implementation of driver terms. All blue lines, and all red lines can be implemented in parallel, respectively. Green lines, which are caused by implicitly enforced three-body constraints, only add a small contribution to the depth and are partially parallelizable with the other steps. In each submodule, one driver line has been omitted as it can be obtained via symmetric difference of the others.}
\label{fig:line-breaker}
\end{figure}

Given a compiled problem layout, i.e.\ the distribution of three- and four-body constraints, we subdivide the entire square lattice into \textit{modules} involving at most ${l_\text{max}\times l_\text{max}}$ qubits, separated by rows and columns of explicitly enforced constraints (see Fig.~\ref{fig:line-breaker}). As a consequence, this limits the length of the driver lines within a module and each module can be treated separately when constructing driver lines. In particular, if all three-body constraints within a module are enforced explicitly, i.e.\ there are only vertical and horizontal lines, the maximal length of a driver line is given by $l_\text{max}$.  Therefore, the circuit depth of the driver Hamiltonian implementation scales linearly with $l_\text{max}$, which is a user-determined and problem-independent quantity that can be chosen in accordance with device-specific needs.
Even in the more general case of conserving some of the three-body constraints within a module implicitly, the problem of finding appropriate hybrid driver lines now reduces to smaller, separate problems for each module and the maximal length of the driver lines will still be approximately $l_\text{max}$.
In any case, the circuits implementing the respective driver terms for each module can be executed simultaneously. 
Considering that $e^{-i\gamma\hat H_\text{Z}}$ and $e^{-i\Omega\hat H^{\rm hyb}_\text{C}}$ can also be implemented with constant-depth circuits, we see that a single QAOA cycle [see Eq.~\eqref{eq:qaoa_protocol_hybrid}] for arbitrary problem-sizes can be implemented with a constant circuit depth.

\section{Initial state preparation}\label{sec:initial_state_preparation}
In order to obtain a suitable initial state $\ket{\psi_0}$ for the QAOA protocol described in Eq.~\eqref{eq:logical_qaoa_protocol} we want to prepare the system in an equal superposition of all computational states spanning $\mathcal{H}_\text{hyb}$, which includes the limiting cases $\mathcal{H}_\text{phys}$ and $\mathcal{H}_\text{CF}$. 
Consider the hybrid driver Hamiltonian $\hat H_\text{X}^\text{hyb}$ involving a valid set of driver lines $D$.
The desired state $\ket{\psi_0}$ is the simultaneous eigenstate of all driver terms in $\hat H_\text{X}^\text{hyb}$ and all implicitly preserved constraints, with eigenvalues $+1$ and $0$ respectively. While this can be easily achieved in the purely explicit approach by preparing each physical qubit in the $\ket{+}$-state, the initial state preparation is more challenging in the implicit and especially the hybrid approach. 
There are known methods to construct circuits generating such a stabilizer state from a trivial product state \cite{Aaronson2004, Garcia2012}, however, the resulting circuits are not necessarily straightforward to implement and might result in large circuit depths on architectures with limited connectivity.

In the following, we propose a simple initialization procedure with a low circuit depth. Resembling the concept of relating a constraint-preserving driver line to a logical qubit (cf. Sec.~\ref{sec:implicit_parity_qaoa}) we now introduce a conceptual \textit{driver qubit} for each hybrid driver line, such that the associated driver term $\hat X^{(\mu)}$ acts as the bit-flip operator on that driver qubit. These ${|D| = \log_2 \dim(\mathcal{H}_\text{hyb})}$ driver qubits represent the states of the considered Hilbert space (satisfying all implicitly enforced constraints).

Thus, the desired initial state corresponds to all driver qubits being in the $\ket{+}$-state, which can be obtained from the $\ket{\uparrow}$-state through consecutive rotations around the $x$- and $z$-axis.
To this end, we also define the phase-flip operator $\hat Z^{(\mu)}$ acting on driver qubit $\mu$, analogous to the stabilizer formalism introduced in Ref. \cite{Rocchetto2016_stabilizer}. The newly defined operators must fulfill the Pauli commutation relations
\begin{equation}
    \{\hat X^{(\mu)}, \hat Z^{(\mu)}\}=[\hat X^{(\mu)}, \hat Z^{(\nu)}]=0
\end{equation}
for ${\mu\neq \nu}$.
For a single driver line $Q^\mu$, it is easy to show that any operator $\hat\sigma_z^{(k)}$, acting on a physical qubit ${k\in Q^\mu}$, fulfills the desired commutation relations with the corresponding $\hat X^{(\mu)}$-rotation. 
As long as this qubit is not involved in any other driver lines, this remains a valid choice.
For example, in the fully implicit implementation shown in Fig.~\ref{fig:all_to_all}b, it is possible to do the $\hat\sigma_z$-rotations on the physical qubits involving the index 0, since each of them is only involved in a single driver line. 

If none of the physical qubits involved in a driver line $Q^\nu$ is exclusively part of this line, this construction fails, as any possible $\hat\sigma_z$-rotation will introduce crosstalk to other driver qubits\footnote{An operator $\hat\sigma_z^{(k)}$ on a physical qubit $k$ translates to the product of $\hat Z$-operators acting on all driver qubits whose driver lines include the physical qubit $k$.}.
However, this rotation does not affect the other driver qubits if they are still in a $\hat{Z}$-eigenstate.
Thus, with an appropriate order of rotations on the driver qubits, it is still possible to use the same state preparation protocol. Additional details for arbitrary driver configurations can be found in Appendix~\ref{sec:appendix:initial_state_preparation}.

Having defined the bit- and phase-flip operators for the driver qubits we can now prepare all driver qubits in the $\ket{+}$-state. We start with the constraint-fulfilling state $\ket{\uparrow}^{\otimes K}$ (in the basis of physical qubits), which corresponds to all driver qubits being in the $\ket{\uparrow}$-state as well. To prepare $\ket{\psi_0}$ from this state, we have to perform physical operations corresponding to a $\pi/2$-rotation around the y-axis on all driver qubits. These operations can be decomposed into consecutive rotations $e^{-i \frac{\pi}{4} \hat X^{(\mu)}}$ and $e^{-i \frac{\pi}{4} \hat Z^{(\mu)}}$, and thus be implemented with the previously defined operators. The circuit depth of the resulting initialization procedure scales the same as the implementation of $\exp(-i \beta \hat H_{X}^\text{hyb})$.

\section{Numerical Results}
\subsection{Circuit depth scaling}
\begin{figure}[tb]
    \centering
    \includegraphics[width=0.9\columnwidth]{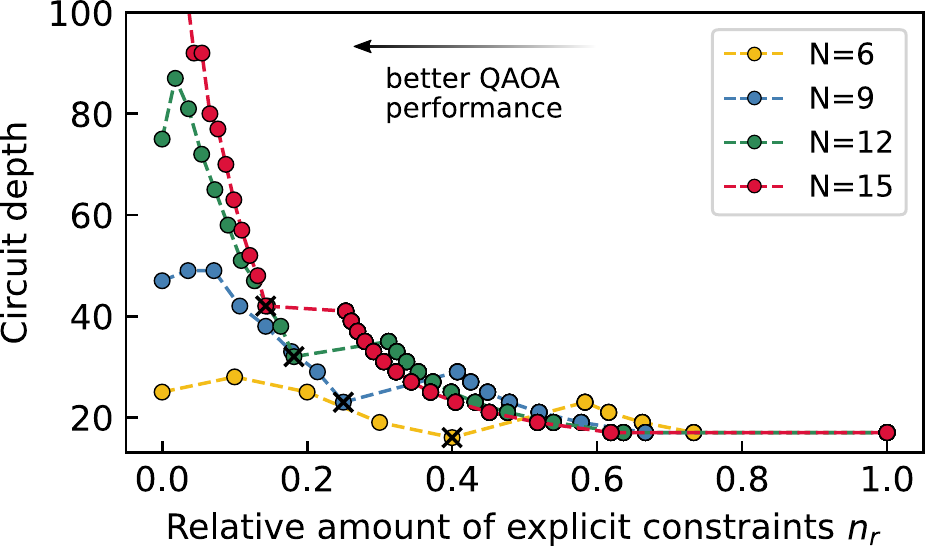}
    \caption{Circuit depth required to implement one step in the QAOA protocol for complete graphs of various problem sizes $N$. The $x$-axis represents different divisions of the constraints into explicit and implicit implementations: For small values of $n_r$, only three-body constraints are enforced explicitly. The points marked with crosses represent implementations where all three-body but no other constraints are enforced explicitly. Larger values of $n_r$ correspond to cases where module-dividing four-body constraints are enforced explicitly as well.
    }
    \label{fig:lhz_depth}
\end{figure}

Figure~\ref{fig:lhz_depth} shows the required circuit depth to implement a single step of the QAOA protocol for a complete problem graph (see Fig.~\ref{fig:all_to_all}) as a function of the relative amount of explicitly enforced constraints ${n_\text{r}=n_\text{C}/n_\text{C}^{\text{tot}}}$. In the fully implicit case (${n_\text{r}=0}$, see Fig.~\ref{fig:all_to_all}b) the circuit depth grows linearly with the system size and the large prefactor in the circuit depth scaling is due to excessive overlap of driver lines. The circuit depth can be reduced by increasing the number of explicitly enforced three-body constraints until, at the points marked by crosses in Fig.~\ref{fig:lhz_depth}, all three-body constraints are explicitly enforced (see Fig.~\ref{fig:all_to_all}c). In this situation the circuit depth still scales linearly with the system size, but with a more favorable prefactor.
All points from there on correspond to modularized layouts of decreasing module size $l_\text{max}$, further improving the circuit depth. Initially this can lead to a small increase of the circuit depth due to the additional implementation cost of the constraint Hamiltonian. This depth-increase is independent of the system size as all explicitly enforced constraints can be implemented in parallel. 
For sufficiently large lattices, the relation between reachable circuit depth and relative amount of explicitly enforced constraints becomes independent of the system size.
The points at ${n_r=1}$ correspond to the fully explicit implementation (see Fig.~\ref{fig:all_to_all}a).

\subsection{QAOA performance}
In order to demonstrate the advantages of this new approach, we compare the QAOA performance of the fully implicit ($n_r=0$), the hybrid ($0<n_r<1$) and the fully explicit ($n_r=1$) parity-QAOA protocol. Subsequent to noiseless QAOA simulations to find optimal parameters, we simulate the resulting optimal QAOA circuits (including the respective state-preparation circuit) under varying noise levels of the required CNOT-gates using qiskit~\cite{qiskit}. The single-qubit gate error rate is kept constant at $10^{-3}$. The simulations were done for various problem instances of a complete graph with ${N=6}$, corresponding to ${K=15}$ physical qubits (see Fig.~\ref{fig:all_to_all}).

\begin{figure}[tb]
    \centering
    \includegraphics[width=0.8\columnwidth]{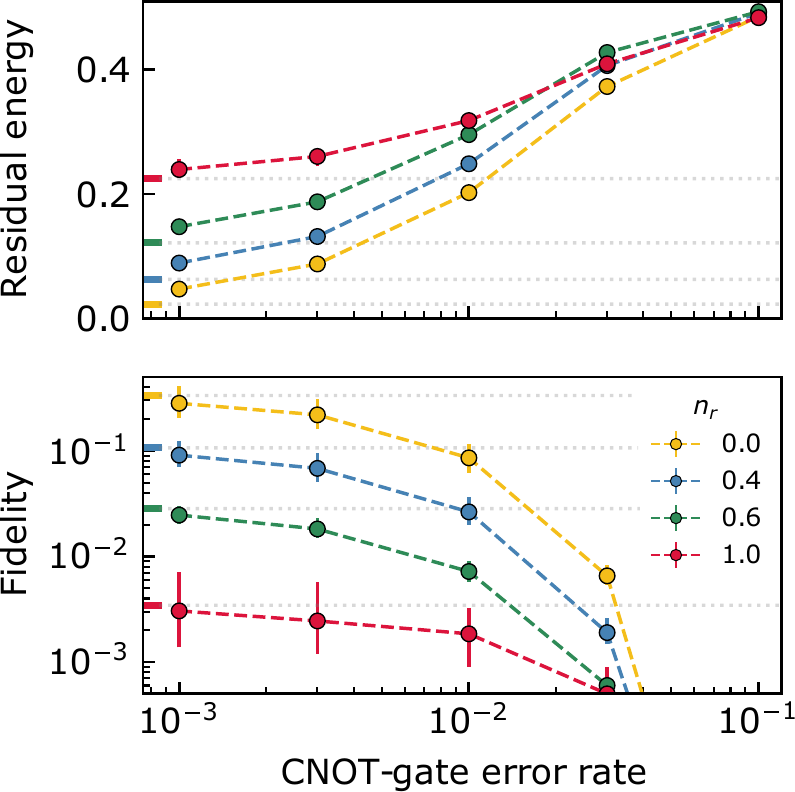}
    \caption{QAOA performance as a function of the CNOT-gate error rate for various relative amounts of explicitly enforced constraints $n_r$. Median residual energy and fidelity were averaged over 96 problem instances of complete graphs with $N=6$ nodes, corresponding to $K=15$ physical qubits. The colored markers on the y-axis indicate the median residual energy/fidelity for the noiseless case.
    The errorbars represent the 25th and 75th percentile over the problem instances.
    }
    \label{fig:qaoa_performance_vs_noise}
\end{figure}

Each data point in Fig.~\ref{fig:qaoa_performance_vs_noise} represents the median performance of 96 problem instances for complete graphs with random local fields $J_m$ [cf. Eq.~\eqref{eq:localfield_hamiltonian}], drawn from a uniform distribution $\mathcal{U}_{[-1, 1]}$. 
For the noiseless parameter optimization of each problem instance, we repeatedly initialize the QAOA-parameters for ${p=3}$, equally distributed in the range ${[0, 2\pi)}$, and search for a local optimum of the energy expectation value. 
Note that for the fully implicit approach there is one QAOA-parameter less per cycle as the constraint unitary has been removed. 
For each initialization we perform consecutive updates of random QAOA-parameters until the energy expectation value converges to a local minimum.
If the energy of the system decreases after a parameter update, the new parameter is accepted, otherwise rejected. 
After repeating the initialization and optimization 100 times the lowest energy expectation value ${E=\braket{\psi | \hat H_\text{phys} | \psi}}$ [cf.\ Eq.~\eqref{eq:qaoa_state_lhz}] for each instance is kept.
We subsequently calculate the residual energy $E_\text{res}$ of the system, defined as
\begin{equation}\label{eq:res_energy}
    E_\text{res}=\frac{E-E_\text{min}}{E_\text{max}-E_\text{min}}\text{,}
\end{equation} 
where $E_\text{max}$ and $E_\text{min}$ denote the maximal and minimal eigenvalues of $\hat H_\text{phys}$. Furthermore, the corresponding fidelity is given by the ground state population with respect to $H_\text{phys}$.

Clearly, the QAOA performance increases with decreasing number of explicitly enforced constraints $n_r$. 
This is related to the fact that with increasing $n_r$, the search space grows and thus additional terms complicate the cost function to be minimized. 
Moreover, we observe that up to error rates of about $10^{-2}$, the different approaches show similar noise-dependency with respect to the QAOA performance. Note that for all values of $n_r$ the absolute CNOT-gate counts are similar and show the same $N$-scaling.

\section{Conclusion and Outlook}
In summary, we have shown how to improve the parity-QAOA performance by interpolating between the standard single-qubit driver Hamiltonian and a driver Hamiltonian tailored to the parity architecture. In particular, the proposed hybrid approach keeps the parallelizability of the fully explicit parity QAOA while gaining performance by reducing the overall search space.
As the key-point of our approach, the trade-off between circuit depth and QAOA performance can be dynamically chosen according to hardware-specific needs by adjusting the size of the implicitly driven submodules.
The presented ideas can be readily realized on any hardware platform providing a regular grid of qubits connected via nearest-neighbor gate operations. This is crucial for addressing questions about the practical QAOA performance of modularized layouts for problem sizes inaccessible to classical simulations.

While the present work focuses on improving the quantum implementation part of the parity QAOA, there are additional opportunities to improve its performance via the classical part, for example different decoding strategies that result in smarter cost functions. 
More generally, further improvements of the parity QAOA might also involve exploiting recently investigated phenomena regarding QAOA parameters \cite{Streif2020, Akshay2021, Wurtz2021} and utilizing other types of mixing Hamiltonians \cite{Govia2021}.

\textit{Acknowledgements - } We thank C. Ertler, G. B. Mbeng, B. E. Niehoff and L. Stenzel for valuable comments and discussions. Work at the University of Innsbruck is supported by the European Union program Horizon 2020 under Grants Agreement No.~817482 (PASQuanS), and by the Austrian Science Fund (FWF) through a START grant under Project No. Y1067-N27 and the SFB BeyondC Project No. F7108-N38.

%\bibliography{references.bib}

\appendix
\onecolumngrid
%~
%\vspace{0.05cm}
%\twocolumngrid
\section{Decomposition of driver terms}
\label{sec:app_decompose}
\begin{figure}[bt]
    \centering
    \includegraphics[width=0.4\columnwidth]{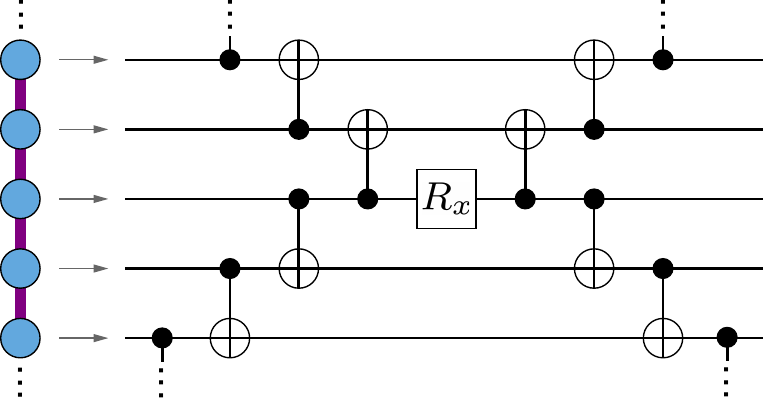}
    \caption{Circuit diagram for decomposing the operator $e^{-i\beta \hat X^{(\mu)}}$ associated with a driver line into CNOT- and $R_x$-rotation gates. The filled small circles denote the control qubit.
    }
    \label{fig:line_circuit}
\end{figure}

Fig.~\ref{fig:line_circuit} shows a possible decomposition of the unitary corresponding to the evolution under a driver term [cf.\ Eq.~\eqref{eq:logical_operators}] into CNOT gates and rotations. Any driver line containing $n$ connected qubits can be implemented with a circuit depth of at most $n+2$. Note that there are many representations of this operator as a circuit \cite{Cowtan2020}, for example the qubit used for the rotation can be freely chosen. This freedom can be used to minimize the circuit depth of a sequence of such operators.

\section{Instructions for initial state preparation}\label{sec:appendix:initial_state_preparation}
As initial state $\ket{\psi_0}$ for the different QAOA-protocols we want to prepare the superposition of all valid states in the considered Hilbert-space. This state can be prepared from $\ket{\uparrow}^{\otimes K}$ via rotations around the y-axis on all driver qubits.
In the fully physical case the required operations trivially translate to single-qubit rotations on all physical qubits, while in the fully implicit and especially the hybrid case these operations involve multiple overlapping driver lines and the order of implementation has to be considered to achieve a valid and low-depth state preparation circuit.

The case where a physical qubit is involved in multiple driver lines has to be treated with care.
Implementing a physical $\hat\sigma_z$-operation on such a qubit has an effect on all involved driver lines and thus can introduce unwanted crosstalk. Whenever possible, we must therefore choose a qubit which is not involved in any other driver lines to perform the phase operation on. 
If this is not possible, the $\hat Z^{(\mu)}$-operation, performed on a qubit $k$ for a driver line ${Q^\mu \ni k}$ can still be used, as long as all driver qubits associated with other driver lines $Q^\nu$ involving qubit $k$ are in an eigenstate of $\hat Z^{(\nu)}$ and thus not affected by the rotation. In the initially prepared state $\ket{\uparrow}^{\otimes K}$, all driver qubits are in the $\hat Z$-eigenstate. That enables us to find a sequence of driver rotations such that for every $\hat Z^{(\mu)}$-rotation there is at least one qubit of the corresponding driver line which is either not included in any other driver lines, or only involved in driver lines whose state has not been rotated yet.
After initializing all $K$ physical qubits in $\ket{\uparrow}$ we assign a priority to every driver line, such that $\hat X$- and $\hat Z$-rotations 
\begin{equation}
    R_\lambda = e^{-i\frac{\pi}{4}\hat Z^{(\lambda)}}e^{-i\frac{\pi}{4}\hat X^{(\lambda)}}
\end{equation}
on the driver qubits, applied in descending order in their priority, will transform all driver qubits into the $\ket{+}$ state.
Here $\lambda$ enumerates the driver lines and the $X^{(\lambda)}$/$Z^{(\lambda)}$ describe bit-flip/phase-flip operations on the corresponding driver qubit.

The priorities of the lines can be found iteratively, we call every line ``unassigned'' until it has been assigned a priority:
\begin{enumerate}
    \item Assign all lines $Q_\mu$ which contain at least one qubit which is not in any other lines the priority ${P_\mu=0}$.
    \item Assign all unassigned lines $Q^\nu$ which overlap at least one line with priority $P_\mu$ (and do not overlap other, unassigned lines at the same qubit) the priority ${P_\nu=P_\mu + 1}$
    \item Repeat step 2 until all lines have a priority.
\end{enumerate}
The initial state $\ket{\psi_0}$ can then be prepared as
\begin{equation}
    \ket{\psi_0} = \prod_{\kappa=0}^{P_\text{max}} \prod_{Q^\mu\in D_\kappa} e^{-i \frac{\pi}{4} \hat Z^{(\mu)}} e^{-i \frac{\pi}{4} \hat X^{(\mu)}} \ket{\uparrow}^{\otimes K},
\end{equation}
where $P_\text{max}$ is the highest assigned priority and ${D_\kappa \subseteq D}$ is the subset of driver lines with priority $\kappa$. Note that the order of the products must be such that the terms with higher priority are applied first.
Equal priorities can be implemented in any order, their required gate sequences can be performed in parallel (or as parallel as possible, if there are qubit overlaps of the driver lines).

\begin{figure}[bt]
    \centering
    \includegraphics[width=.3\columnwidth]{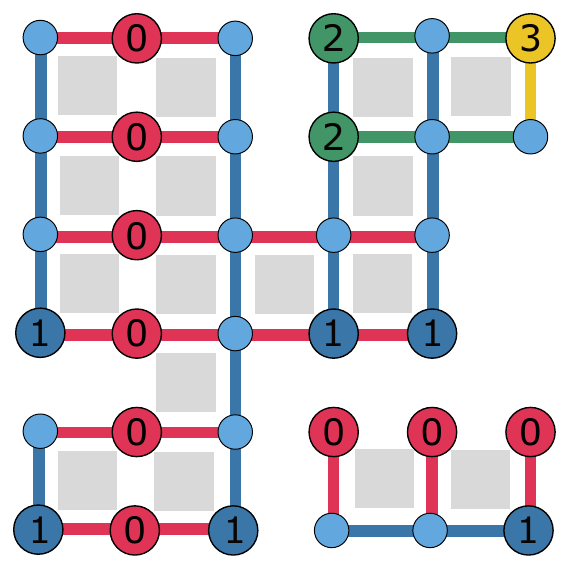}
    \caption{Illustration of the initial state preparation for a module with two sets of connected lines. Only implicitly enforced constraints are shown. The line colors and numbers indicate the priority $P_\mu$ of a line, with the highlighted qubits showing an example of where the physical $\hat \sigma_z$-rotations may be implemented. Rotations on lines with higher priority must be performed first.
    }
    \label{fig:line-priorities}
\end{figure}

If with this procedure, not all lines can be assigned a priority, the partitioning of constraints can be changed to include more explicitly enforced constraints. At the latest, a valid prioritizing of all lines can be found once all three-body constraints are explicitly enforced.
Fig.~\ref{fig:line-priorities} shows an example of two sets of connected driver lines in a sub-module with assigned priorities.

\end{document}